\documentclass[12pt]{article} 
\usepackage[dvips]{graphicx}
\usepackage{amssymb}
\usepackage{amsmath} 
\usepackage[mathcal]{eucal}
\usepackage{latexsym}
 
\begin{document}   
\newcommand{\todo}[1]{{\em \small {#1}}\marginpar{$\Longleftarrow$}}   
\newcommand{\labell}[1]{\label{#1}\qquad_{#1}} %{\label{#1}} %  

\vskip 1cm 
\vskip 1cm 

%%                      Title here  
%%  

\begin{center} 
{\Large \bf The No-Boundary Probability for the Universe starting at the top of the hill.}
\end{center} 
\vskip 1cm   
  
\renewcommand{\thefootnote}{\fnsymbol{footnote}} 
\centerline{\bf T. Clunan\footnote{T.P.Clunan@damtp.cam.ac.uk}}
\vskip .5cm   

\centerline{ \it DAMTP, Centre for Mathematical Sciences, University of Cambridge}   
\centerline{ \it Wilberforce Road, Cambridge. CB3 0WA. United Kingdom.}   
\setcounter{footnote}{0}   

\renewcommand{\thefootnote}{\arabic{footnote}}

%%                      Text starts here  
%% 

\begin{abstract}   
We use the Hartle-Hawking No-Boundary Proposal to make a comparison between the probabilities 
of the universe starting near, and at, the top of a hill in the effective potential.  In 
the context of top-down cosmology, our calculation finds that the universe doesn't start at 
the top.
\end{abstract}

\section{Introduction}
\label{introduction}

It is common to consider the problems of big bang cosmology, such as the horizon 
problem, to be cured by having a period of inflation.  A positive cosmological constant would give 
inflation, but we also need the period of inflation to have a natural 
end.  The simplest way to do this is to have a scalar field slowly rolling in a 
potential, with a minimum in which we sit today. 

One problem with a big bang type initial singularity is that the laws of physics
break down at the singularity, and we therefore know little about the initial
conditions of the universe.  However, a theory of the universe which is only a
theory of evolution can never be complete - it would mean that we would have to guess
at the initial conditions.  In order to have a complete theory we will also need a
theory of the initial state of the universe.  The no boundary proposal of Hartle and
Hawking~\cite{wavefunctionoftheuniverse} is the theory of initial conditions we
shall choose to work with.  

The potential may be very complicated and have numerous valleys and hill tops.  We 
will work with the top-down cosmology approach, and so we constrain the universe to 
start some way off the minimum in which reheating happens, in order that we get a large 
enough density.  In this paper we will consider a hill where the conditions imposed 
through top-down cosmology are such that the universe starts somewhere close to the top of 
the hill; in this case, we will ask whether the universe in fact starts in an unstable 
de-Sitter state at the top of the hill.\footnote{In the case that the universe starts at the top, 
a perturbation will cause roll-down into the Lorentzian region~\cite{nonsingularuniverse},
so the two possibilities may fairly be compared.}
We will consider a broad potential, with\footnote{There 
has been interest in this case before~\cite{whydoesinflationstartatthetopofthehill}} 
$0>V_{,\phi \phi}/V>-4/3$ at the top of the hill, so that we are considering either a Hawking-Moss instanton or a 
Hawking-Turok instanton; Coleman-De Luccia instantons~\cite{colemandelucciainstanton} do not arise in this case. 

The instantons are selected by prescribing appropriate quantities on their boundary.  In 
the past, numerous authors have considered the problem in the minisuperspace by 
prescribing the scalar field $\phi$ and the scale factor $b$ on the boundary of the 
instanton.  Instead we will prescribe $\phi$ and the trace of the second fundamental form
$K$ on the boundary of the instanton.  Our approach is in keeping with top down 
cosmology~\cite{topdown} and the expansion of the universe.

There are many papers which consider instantons which are composed of a real Euclidean section, and a real Lorentzian 
section, analytically joined together; and there are a number of authors who consider fully complex instantons.  
Here we will consider fully complex instantons.  In order to calculate the probability of 
a given instanton, we will need to calculate its action.  We will be comparing the actions 
for the Hawking Moss instanton with the Hawking Turok instanton.  Since we are only 
considering a comparison for Hawking Turok instantons close to the top of the hill, we can 
address the question with a perturbation calculation; and we expect to have to calculate
the action of the Hawking Turok instanton to second order in the perturbation
parameter\footnote{This problem has been addressed previously 
in~\cite{whydoesinflationstartatthetopofthehill}.  There the case $0>V_{,\phi \phi}/V>-4/3$ was also
considered and, as in this work, the solution for the scale factor and scalar field were 
computed in an expansion in the initial displacement of the scalar field from the potential maximum.
There however, the scale factor was not consistently corrected to second order; this we shall rectify 
here.  The resulting extra contributions to the action are responsible for the difference of our 
result from theirs.}. 
This is what we proceed to do in the next section.  We will also see that is is sufficient to perform a 
second order perturbation calculation.

\section{Finding the action}
The Euclidean action for a real scalar field plus gravity is
\begin{equation}
S=-\frac{1}{2\kappa}\int_{M} d^{4}x \sqrt{g} R -\frac{1}{3 \kappa}\int_{\Sigma} d^{3}x \sqrt{h} K 
+\int_{M} d^{4}x \sqrt{g} \left( \frac{1}{2} g^{\mu\nu} \partial_{\mu} \phi \partial_{\nu} 
\phi + V(\phi)\right),
\end{equation}
where $\kappa = 8 \pi G$, $M$ is the no-boundary manifold and $\Sigma = \partial M$ is 
its boundary (a copy of $S^{3}$).  The boundary term is chosen to be appropriate to the variational problem
where $K$, $\phi$ and $\overline{h_{ij}}$ are held fixed on the boundary (here $\overline{h_{ij}}=\gamma_{ij}$ 
is the metric on a unit $S^{3}$).  This boundary term is different from the case where $b$ is fixed on the 
boundary.  Using the resulting equation of motion, $R=\kappa ((\partial\phi)^{2}+4 V(\phi))$, this becomes
\begin{equation}\label{general_action}
S=-\frac{1}{3 \kappa}\int_{\Sigma} d^{3}x \sqrt{h} K -\int_{M} d^{4}x \sqrt{g}  V(\phi).
\end{equation}
We will take the potential to be
\begin{equation}\label{potential}
V (\phi) = A (1-C \phi^{2})
\end{equation}
with $A>0$, $C>0$ because we are interested in considering a hill top with 
positive potential\footnote{It has been argued~\cite{fastrollinflation} that in the Lorentzian 
region the homogeneous scalar field cannot relax in a negative minimum, so the present day 
potential minimum in which the universe sits must be non-negative, and so certainly any hill 
top at which it starts must have a positive potential.}; and $C<2/3$ so that the hill top is 
broad enough ($0>V_{,\phi \phi}/V>-4/3$) for us not to get a Coleman-De Luccia 
instanton~\cite{whydoesinflationstartatthetopofthehill,colemandelucciainstanton}.  It should be 
noted that we will only be interested in the region around the top of the hill, so it is 
not a problem that our potential is unbounded below, and nor is it a problem that our 
potential doesn't feature a minimum.

\subsection{Perturbation calculation}
We consider the homogeneous $O(4)$ symmetric case\footnote{As
in~\cite{wavefunctionoftheuniverse,whydoesinflationstartatthetopofthehill} we restrict attention to continuation to 
closed universes.} where the metric in the Euclidean region is 
\begin{equation}
ds^{2}= d\tau^{2}+b^{2}(\tau) d\Omega_{3}^{2}
\end{equation}
and $\phi=\phi(\tau)$.  In particular we consider the homogeneous perturbation from the 
Hawking-Moss instanton.  We take the perturbation parameter to be the value of $\phi$ at
the regular pole of the Hawking-Turok instanton: $\epsilon= \phi (0)$. Then
\begin{eqnarray}\label{pertseriesforphi}
\phi & = & \phi_{0}+\epsilon \phi_{1}+ \epsilon^{2} \phi_{2} +\ldots , \\ \label{pertseriesforb}
b & = & b_{0}+\epsilon b_{1} + \epsilon^{2} b_{2}+\ldots
\end{eqnarray}
with $\phi_{0} \equiv 0$ and $b_{0} ( \tau ) = h^{-1} \sin ( h \tau )$, where 
$h=\sqrt{A/3}$. The conditions for $\tau=0$ to be the regular pole of the instanton are 
\begin{equation}
\phi (0)=\epsilon,\;  \phi '(0)=0,\; b(0)=0, \; b'(0)=1.
\end{equation}

Setting $\kappa=1$, the classical constraint equation is 
\begin{equation}\label{constraint}
\left(\frac{b'}{b}\right)^{2} = \frac{1}{3} \left( \frac{1}{2} (\phi ')^{2} -V \right) +\frac{1}{b^{2}}\, ;
\end{equation}
and the classical equations of motion for the Euclidean metric are 
\begin{eqnarray}
0=b \phi ''+3 b' \phi ' -b V_{,\phi} \, ,\;\\
0=b''+\frac{1}{3} b \left( (\phi ')^{2} + V   \right).
\end{eqnarray}  
These tell us that $\phi$ is odd in $\epsilon$ while $b$ is even.  Thus by finding 
$\phi_{1}$ and $b_{2}$ we may calculate the action to second order in $\epsilon$ (and since the question is
whether there is a local maximum in the probability for nucleation at the top of the hill, this is all that is
necessary). The solution is
\begin{equation}\label{firstorderphi}
\phi_{1}(\tau)=\;F_{\!\!\!\!\!\!\!\!2\;\;\;\;1} \Big(\frac{3}{2}+q,\frac{3}{2}-q,2,z(\tau)\Big)\, ,
\end{equation}
where $z(\tau)=\sin^{2}(\frac{h \tau}{2})$ and $q=\sqrt{9/4+6C}$; and
\begin{equation}
\label{problemintegral}
b_{2}(\tau)=-\int_{\lambda=0}^{\tau} \sin (h(\tau - \lambda)) \sin (h \lambda) \left( \frac{1}{3 h^{2}}
(\phi_{1}^{'}(\lambda))^{2} - C \phi_{1}^{2}(\lambda)\right) d \lambda \, .
\end{equation}
We can approximate (\ref{problemintegral}) by approximating $\phi_{1}(\tau)$ by a truncated series in $\tau$.
Eg.
\begin{equation}
\phi_{1}\approx 1-\frac{3}{4} c h^{2} \tau^{2}
\end{equation}
in (\ref{problemintegral}) would give
\begin{eqnarray}\nonumber
b_{2}(\tau) \approx & \frac{-c}{320 h} \big( [(160+180 c +135 c^{2})-30 c (4+3 c) h^{2} \tau^{2} + 18 c^{2} h^{4} \tau^{4}] h \tau cos(h \tau) \\
\nonumber
& -5 [32 +36 c + 27 c^{2} - 9 c (4+3 c) h^{2} \tau^{2} + 9 c^{2} h^{4} \tau^{4}] sin (h \tau) \big)
\end{eqnarray}

\noindent From the expression (\ref{general_action}) for the action we have
\begin{equation}\label{action_in_terms_of_b}
S=-4 \pi^{2} \int_{\tau =0}^{\tau_{\Sigma}} b d \tau \, .
\end{equation}

\noindent We are using $K(\tau)=3 b'(\tau)/ b(\tau)$ and $K_{0}(\tau)=3 b_{0}'(\tau)/ b_{0}(\tau)$. 
This gives
\begin{equation}\label{kequation}
K=K_{0}+\frac{\epsilon^{2}}{b_{0}} \left( 3 b_{2}' - K_{0} b_{2}  \right)+\ldots .
\end{equation}
We define $\tau_{0}$ to be such that $K_{0}(\tau_{0})=K_{\Sigma}\equiv K(\tau_{\Sigma})$;
then $\tau_{\Sigma}=\tau_{0}+\delta \tau$ with (from (\ref{kequation})):
\begin{equation}
\label{deltatau} \delta \tau = -\frac{\epsilon^{2}}{K_{0}'(\tau_{0})
b_{0}(\tau_{0})}\left(3 b_{2}'(\tau_{0})-K_{\Sigma} b_{2}(\tau_{0})\right)+\ldots .
\end{equation}

\noindent From (\ref{action_in_terms_of_b}) we have	
\begin{equation}\label{actionwithdeltatau}
\delta S = - 4 \pi^{2} 
\bigg(\frac{\phi_{\Sigma}}
{\phi_{1}(\tau_{\Sigma})}\bigg)^{2} \Big(\frac{- b_{0}(\tau_{0}) \left(3 b_{2}'(\tau_{0})-K_{\Sigma}
b_{2}(\tau_{0})\right)}{K_{0}'(\tau_{0})b_{0}(\tau_{0})} + \int_{0}^{\tau_{0}}  b_{2} d \tau \Big)+\ldots ,
\end{equation}
where we have used 
\begin{equation}\label{attheboundary}
\phi_{\Sigma}=\epsilon \phi_{1}(\tau_{\Sigma})+\ldots
\end{equation}
to give the series in terms of $\phi_{\Sigma}$.

\subsection{Upper bound on $K_{\Sigma}$}
First we consider the action for the Hawking-Moss instanton: from
(\ref{action_in_terms_of_b}) we see this is
\begin{eqnarray}
S_{HM}= 4 \pi^{2} h^{-2} \Bigg(\frac{K_{\Sigma}}{3 h \sqrt{1+\big(\frac{K_{\Sigma}}{3 h}\big)^{2}}}-1 \Bigg);
\end{eqnarray}
we see that this has branch cuts extending from $K_{0}=\pm 3 \imath h$ along the imaginary Euclidean $K_{\Sigma}$ axis 
away from the origin - thus it doesn't make any sense to try to 
extend from the real line beyond these values when we are considering the case of the Hawking-Moss instanton.  It is
easy to see the cause of this problem:  consider equation (\ref{constraint}); analytically continuing this to
Lorentzian time shows that Lorentzian $K_{\Sigma}$ is bounded by $3 h$.  Considering this same equation we see 
that for Hawking Turok instantons near the top of the hill with roughly constant $\phi$ throughout (these are the
instantons arising in the perturbation calculation) we have a small upper bound on $(\phi^{'})^{2}$ so the upper
bound on Lorentzian $K_{\Sigma}$ only moves at second order in a perturbative expansion in $\epsilon$.  Thus we can 
expect that the branch cuts in the action don't move at first order.  Also, if we wish to compare Hawking Turok
instantons near the top of the hill with Hawking Moss instantons at the top of the hill then we can only consider
the range of $K_{\Sigma}$ applicable to both: i.e. Lorentzian $K_{\Sigma}$ with $0\le K_{\Sigma}\lesssim 3 h$.

\subsection{Numerical calculation}\label{numericalsection}
Fixing $A$ and $C$ and performing the calculations one can show that the results for $\delta S$ are independent of the 
truncation at a given $K_{\Sigma}$ if the order at which we truncate is high enough.  However, for any choice of the power
at which we truncate $\phi_{1}(\tau)$, the results will go wrong for large enough Lorentzian $K_{\Sigma}$.  Thus we chose
to use a numerical method.

We chose a real Lorentzian $K_{\Sigma}>0$ and real $\phi_{\Sigma}$; then we used the perturbation 
calculation above (including the truncation of $\phi_{1}(\tau )$) to estimate values for $\tau_{\Sigma}$ and $\phi (0)$ (using
equation (\ref{deltatau}) and $\phi_{\Sigma}=\phi(0)\;\;F_{\!\!\!\!\!\!\!\!2\;\;\;\;1}
(3/2+q,3/2-q,2,\sin^{2}( h \tau_{\Sigma}/2)) +\ldots$
% $\phi_{\Sigma}=\phi(0) \phi_{1}(\tau_{\Sigma})+\ldots$
).  Both of
these are expected to be complex.  Next, set $\theta=\arg (\tau_{\Sigma})$ and $\tau = \sigma e^{\imath \theta}$ with 
$\sigma \ge 0$ and numerically integrate the equations 
\begin{eqnarray}
0=b \ddot{\phi}+3 \dot{b} \dot{\phi} - e^{2 \imath \theta} b V_{,\phi}\\
0=\ddot{b}+\frac{1}{3} b \left( \dot{\phi}^{2} + e^{2 \imath \theta} V   \right)
\end{eqnarray}
with the boundary conditions $\phi (0) =\phi_{0}$, $\dot{\phi} (0)=0$, $b(0)=0$ and $\dot{b} (0) = e^{\imath \theta}$ 
along real $\sigma > 0$ until we get 
$3 \imath \dot{b}/(e^{\imath \theta} b)$ to be purely real (it is expected that the value of this quantity 
will then be 
close to the original value of $K_{\Sigma}$ selected). Here $\dot{\phi}$ is differentiation of $\phi$ with respect to $\sigma$, etc. 
Calculating the value of $\phi$ at this point we will find 
it has a small imaginary value; by making small changes to the value of $\theta$ and integrating the equations repeatedly we can 
arrive at complex values for $\tau_{\Sigma}$ and $\phi_{0}$ which result in an instanton with real Lorentzian 
$K_{\Sigma}$ and real $\phi_{\Sigma}$ close to the values we originally chose.  Next we compute the action for this 
instanton and the action for the Hawking Moss instanton with this value of Lorentzian $K_{\Sigma}$; we difference 
these and divide by $\phi_{\Sigma}^{2}$ to get an estimate for 
$\lim_{\phi_{\Sigma} \rightarrow 0}(\Re(\delta S)/\phi_{\Sigma}^{2})$, 
which we can compare to our previous calculation.  This is how 
we obtained the plots in Figures \ref{graph1}, \ref{graph2} and \ref{graph3}. 

\subsubsection{Accuracy of numerical estimate for $\lim_{\phi_{\Sigma} \rightarrow 0}(\frac{\Re
(\delta S)}{\phi_{\Sigma}^{2}})$} 
\begin{figure}
\centering
\includegraphics[]{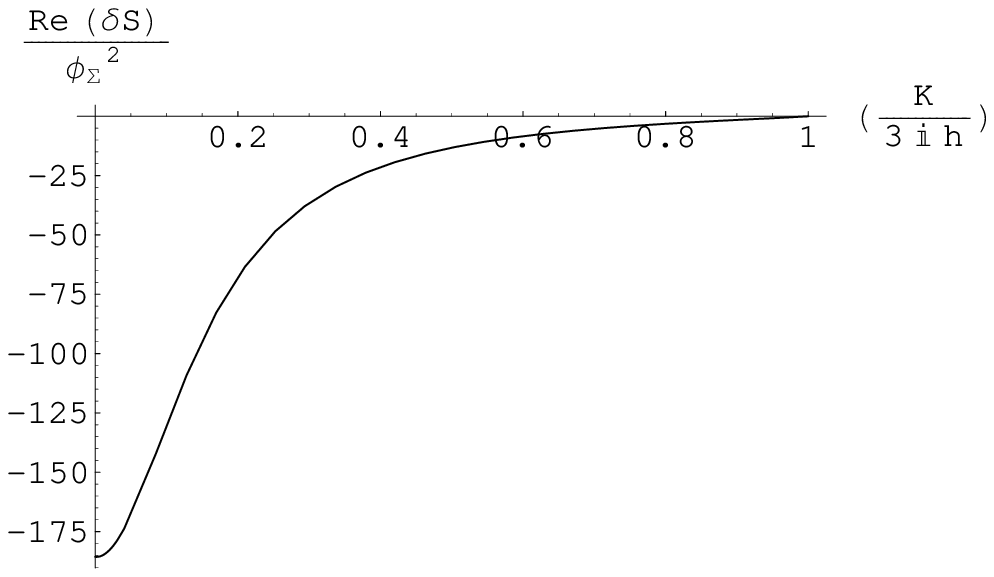}
\caption{With $A=1$ and $C=1/2$ and $\delta S =S_{Hawking-Turok}-S_{Hawking-Moss}$: we see that the universe doesn't start 
at the top of the hill for any of the possible values of real Lorentzian $K_{\Sigma}$.\label{graph1}}
\end{figure}

\begin{figure}
\centering
\includegraphics[]{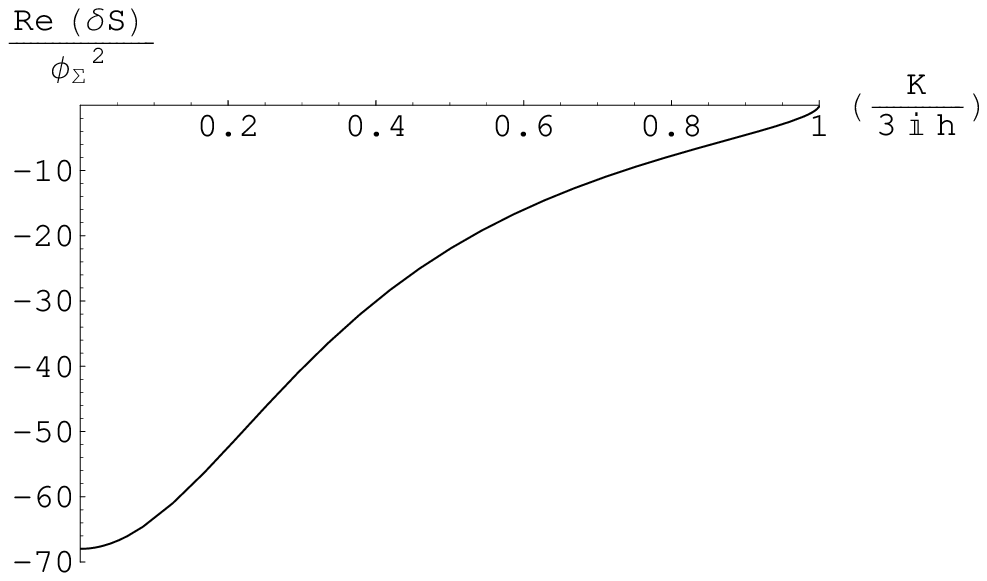}
\caption{With $A=1$ and $C=1/3$ and $\delta S =S_{Hawking-Turok}-S_{Hawking-Moss}$: we see that the universe doesn't start 
at the top of the hill for any of the possible values of real Lorentzian $K_{\Sigma}$.\label{graph2}}
\end{figure}

\begin{figure}
\centering
\includegraphics[]{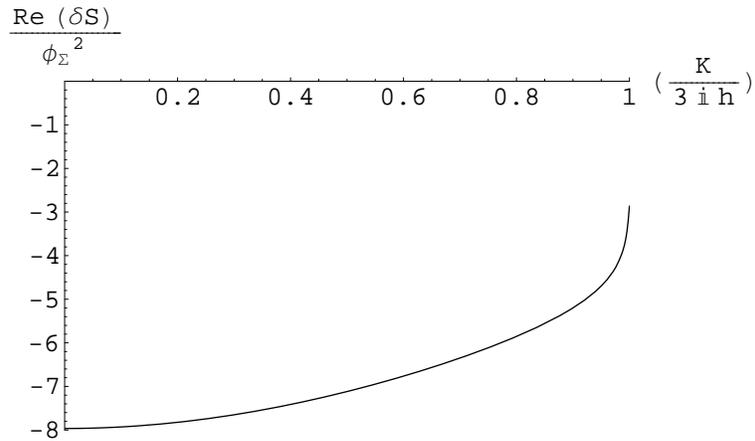}
\caption{With $A=1$ and $C=1/16$ and $\delta S =S_{Hawking-Turok}-S_{Hawking-Moss}$: we see that the universe doesn't start 
at the top of the hill for any of the possible values of real Lorentzian $K_{\Sigma}$.\label{graph3}}
\end{figure}

It is clear that the points we get 
through the process described in section \ref{numericalsection} can have differing values of 
$\Re (\delta S)/\phi_{\Sigma}^{2}$ depending on $\phi_{\Sigma}$ for the instanton under 
consideration.  However, because $\delta S$ can be written as a perturbation series in 
$\phi_{\Sigma}$ with lowest term of order $\phi_{\Sigma}^{2}$, we expect the error in our 
estimate of $\lim_{\phi_{\Sigma} \rightarrow 0}(\Re(\delta S)/\phi_{\Sigma}^{2})$ to be of 
order $\phi_{\Sigma}$.  This was verified by taking $\phi_{\Sigma}\approx 0.1$ and 
$\phi_{\Sigma}\approx 0.01$ and comparing the resulting values of 
$\Re (\delta S)/\phi_{\Sigma}^{2}$ for the two instantons with the value of 
$K_{\Sigma}$ of interest.  The difference was shown to be of order $0.1$.  The 
points plotted in Figures \ref{graph1}, \ref{graph2} and \ref{graph3} are for values of 
$\phi_{\Sigma}\approx 0.01$ and can therefore be expected to give errors in the estimate of 
$\lim_{\phi_{\Sigma} \rightarrow 0}(\Re (\delta S)/\phi_{\Sigma}^{2})$ of order $0.01$.  These 
plots therefore demonstrate that $\lim_{\phi_{\Sigma} \rightarrow 0}(\Re (\delta S)/\phi_{\Sigma}^{2})<0$ 
for all values of Lorentzian $K_{\Sigma}$ and suggest that it tends to zero as Lorentzian 
$K_{\Sigma}$ tends to $3 h$ from below.

\section{Discussion}

Up to a prefactor, the probability of an instanton is given by 
$\exp (-\Re (\textrm{Euclidean Action}))$.  As we may neglect the prefactor for small 
$\phi$, ~\cite{whydoesinflationstartatthetopofthehill}, we compare probabilities by 
considering 
\begin{equation}
\frac{P(\textrm{HT})}{P(\textrm{HM})}=\exp (-\Re (\delta S));
\end{equation}
{\it i.e.} the universe starts at the top of the hill if and only if $\Re (\delta S)>0$.  Looking at the plots in Figures 
\ref{graph1}, \ref{graph2} and \ref{graph3} we conclude that the universe doesn't start at the top of the hill for any of
the possible values of $K/(3 \imath h)$, where $K$ is the trace of the Euclidean second fundamental form, so the universe 
doesn't start at the top of the hill for any of the possible values of real Lorentzian $K_{\Sigma}$.

\section{Conclusion}
By considering the dominant contribution to the path integral appropriate to the top-down 
approach, we have seen that the universe does not start at the top of the hill. 
Future work might be to extend this to a calculation which doesn't just look at the top of 
the hill, so that we can see where the universe does start.

\medskip
\centerline{\bf Acknowledgements}
\medskip    

I wish to thank Stephen Hawking for discusions about this work and for 
prompting interest in it; and also Neil Turok, Paul Davis, Thomas Hertog and Malcolm Perry for 
discussions and encouragement.  I am grateful for support from the EPSRC.

%\bibliographystyle{./utphys}
%\bibliography{tim}

\providecommand{\href}[2]{#2}\begingroup\raggedright\endgroup

\end{document}